\documentclass[aps,prm,twocolumn,showpacs,groupedaddress]{revtex4-1}
\usepackage{amssymb}
\usepackage{amsmath}
\usepackage{float}
\usepackage{graphicx}
\usepackage{color}
\usepackage{soul}
\usepackage[pdftex,colorlinks=true,linkcolor=red,citecolor=blue]{hyperref}
 
\pagecolor{white}
\begin{document}

\title[]{Magnetic anisotropy in uranium monosulfide, probed by the magnetic torque measurements}

\author{Narayan Poudel$^1$}
\email{Narayan.Poudel@inl.gov}%
\author{Jason Jeffries$^2$}
\author{Krzysztof Gofryk$^1$}
\email{gofryk@inl.gov}%

\affiliation{$^1$ Idaho National Laboratory, Idaho Falls, Idaho 83415, USA}
\affiliation{$^2$ Lawrence Livermore National Laboratory, Livermore, California 94550, USA}
\date{\today}%

\begin{abstract}
We have studied the magnetic torque in uranium monosulfide (US) single crystals to explore the magnetic anisotropy in this material. Uranium monosulfide crystallizes in cubic, NaCl-type of crystal structure and exhibits the largest magneto-crystalline anisotropy observed in cubic systems. By performing detailed torque measurements we observe a strongly anisotropic behavior in the paramagnetic phase due to crystal defects and quadrupolar pair interactions. Our studies also confirm the presence of a large anisotropy in the ferromagnetic state of the US system with the $<$100$>$, $<$111$>$, and $<$110$>$ directions being hard, easy, and intermediate axis, respectively. Furthermore, the anisotropy in the paramagnetic phase shows similar characteristics to the anisotropy observed in the ferromagnetic phase, as characterized by second and fourth rank susceptibility terms. The similarity of the anisotropic behaviors in paramagnetic and ferromagnetic phases is the consequence of strong magneto-elastic properties in this system, which possibly lead to the rhombohedral structural distortion, not only in the ferromagnetic phase but also in the paramagnetic phase (induced by applied magnetic field).
\end{abstract}


\maketitle

\section{Introduction}

Anisotropic behaviors have been observed in various cubic 4$f$ and 5$f$ electron-based materials \cite{Erdos:83, Freeman:84, Santini:99, Gofryk:14, Jaime:17}. The main source of anisotropy in these materials is magnetocrystalline anisotropy, which involves large orbital magnetic moment, strong spin-orbit coupling, and hybridization of $5f$-electrons with $p,d$-states of ligand ions  \cite{Bloch:31, Brooks:83, Kioussis:93, Sandratskii:18, Masakova:12, Sandratskii:19, Sandratskii:20}. This is  well emphasized in uranium monosulfide, in which the largest magneto-crystalline anisotropy has been observed for cubic systems \cite{Gardner:68, Tillwick:76b}. The energy density $K_1$ for US at 0 K was reported to be $\sim8.5\times10^{8}$ ergs/cm$^3$ from previous torque measurement \cite{Tillwick:76b} and of about $10^{10}$ ergs/cm$^3$ from the thermal neutron scattering \cite{Lander:90}.

Uranium monosulfide is paramagnetic (PM) at room temperature and it undergoes a magnetic phase transition into the ferromagnetic (FM) phase below $T_C\sim$177 K, with the easy axis of magnetization along [111] \cite{Tillwick:76, Tillwick:77}. Furthermore, the FM phase is characterized by a small rhombohedral distortion along [111] direction, \cite{Wedgwood:72, Lander:90, Herrm:06}. The US crystal also undergoes a structural phase transition from cubic to a rhombohedral structure at room temperature and under the pressure of 10 GPa \cite{Gerward:89}. A detailed study of magnetic properties and its structural distortion under pressure shows that the ordering temperature decreases with applied pressure until the ferromagnetism suddenly disappears near the pressure-induced rhombohedral distortion \cite{Jeffries:13}. 

The U-U distance in uranium monosulfide is greater than the Hill limit \cite{Hill:70} that might indicate a localized character of 5f-electrons in this material. That straightforward approach, however, does not work well in the case of US, which has been shown to exhibit a strong itinerant component in its electronic structure \cite{Kioussis:93}. The itineracy is also supported by low values of both the paramagnetic and ferromagnetic moments, specific heat measurements that imply an enhanced electronic part, as well as photoemission spectra that reveal a narrow $5f$-electron peak near the Fermi level \cite{Tillwick:76b, Freeman:85, Schoenes:84, Baer:80, Baer:80b}. Furthermore, the reduction of magnetic moment under pressure might also suggest that orbital moment is predominant in US  and 5$f$ states are fairly delocalized \cite{Brooks:83, Fournier:85a}. The delocalized nature of 5$f$ states gives  stronger hybridization in uranium compounds  compared to that of the 4$f$ states in rare-earth systems. The complex magnetic behavior and magnetic anisotropy was also observed in moderately hybridized 4$f$ cubic systems where the $f$-electrons hybridize with band electrons \cite{Wills:87, Kioussis:91, Busch:67}.

In general, the magnetic torque measurements performed on single crystals are considered as a sensitive probe for studying the magnetic, especially anisotropic properties of solids. This technique has been successfully used to study anisotropic behaviors of superconductors, heavy fermions, and various antiferromagnetic, ferromagnetic, and paramagnetic materials \cite{Xiao:06, Campbell:17, Hong:16, Steenbeck:11, Yang:19, Rathnayaka:07}. Here, we report detailed measurements of the magnetic torque of US single crystals. Our studies confirm the presence of a strongly anisotropic behavior in the PM and FM phase in this cubic 5$f$-electron system. The observed anisotropy in the PM state shows a similar anisotropy behavior as in the magnetic state. We discuss our results in the context of symmetry change in the paramagnetic phase due to crystal defects and magnetic quadrupole interactions.

\begin{figure}[b]
\begin{center}
\includegraphics[angle=0, width=3.2 in]{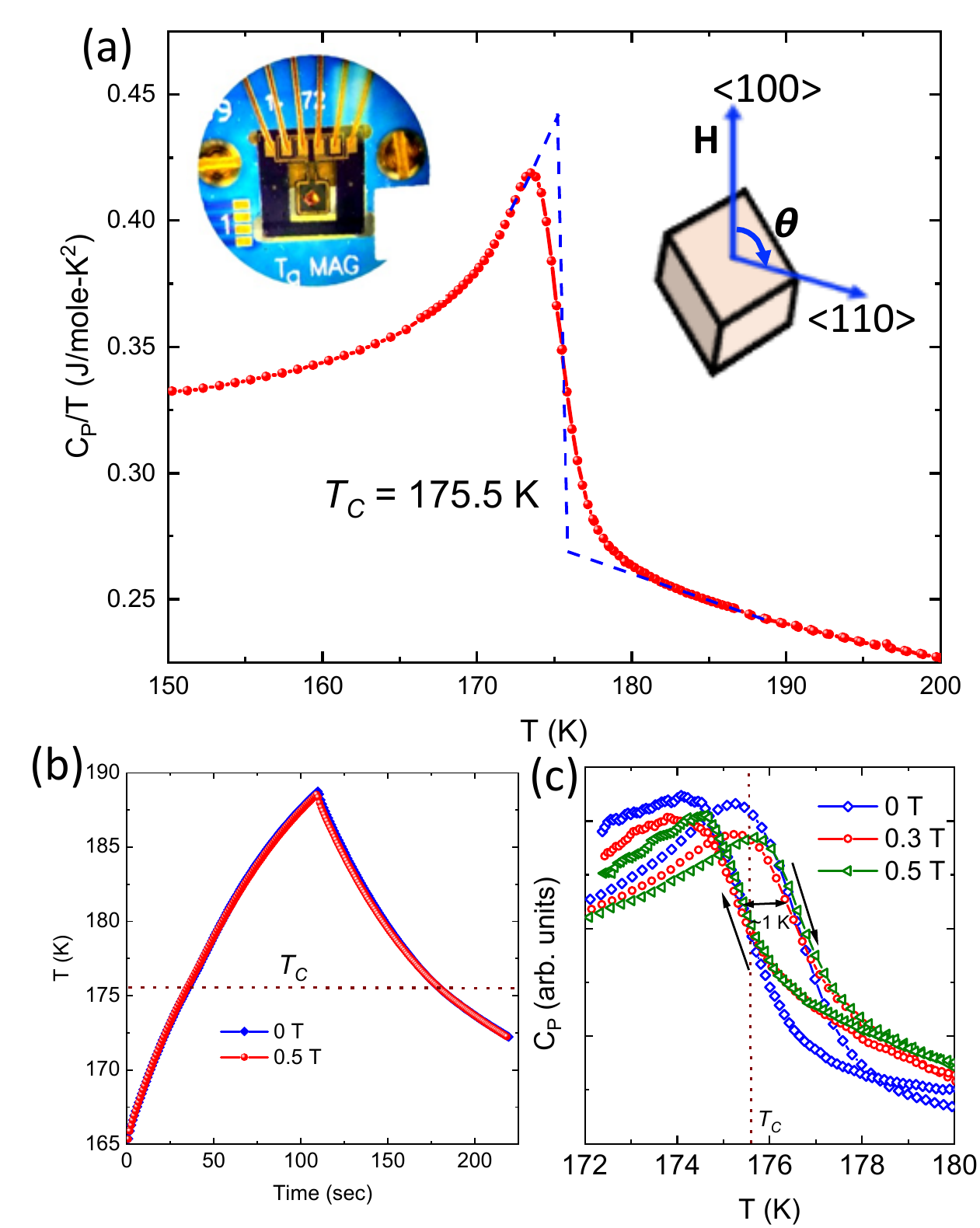}
\end{center}
\caption{(Color online) (a) The temperature dependence of the heat capacity of US crystal in the vicinity of the phase transition. The Curie temperature $T_C$ is found to be 175.5 K. Inset: the sketch of the crystal (as placed on the torque cantilever) and its orientation with respect to rotation axis and direction of applied magnetic field; (b) the heating and cooling curves during the measurements across the magnetic phase transition. As seen, no latent heat anomaly was observed in zero field as well as in magnetic field of 0.5 T. (c) A narrow hysteresis is observed ($\sim$ 1 K) in $C_p(T)$ during heating and cooling the crystal across the phase transition. The tail of the hysteresis is moving towards higher temperature with applied magnetic field, characteristic of ferromagnetic systems.}
\label{fig0}
\end{figure}

\section{Experimental Details}

The magnetic torque of high quality US single crystals was measured using a piezoelectric cantilever and TqMag option implemented in the PPMS Dynacool-9 system (Quantum Design). The torque of a small US crystal (m = 1.08 mg) was measured in wide temperature (2-300 K), magnetic field (0-9 T), and angle (0-360 deg) ranges. The measured crystal was in the shape of a small cube (a $\sim$0.5 mm) to reduce the shape anisotropy effects \cite{Bertotti:98}. The crystal was mounted in such a way that all three principal directions could be probed by rotating the sample in a magnetic field (rotation from $<$100$>$ to $<$110$>$ crystallographic directions). That rotation configuration is equivalent to rotating the magnetic field in $\{$110$\}$ plane by 360$^{\circ}$. In addition to magnetic torque, we also measured the heat capacity of the crystals to determine its Curie temperature $T_C$ (Fig. \ref{fig0} (a)), which is found to be 175.5 K from $C_{P}/T$ versus $T$ graph. As shown in Fig. 1(b), we didn't observe any signatures of latent heat when heating and cooling the sample in the vicinity of $T_C$, even though a small structural distortion (from high-temperature cubic to low temperature rhombohedral) has been reported for this material at Curie temperature \cite{Marples:70}. This might indicate that the rhombohedral distortion along $<$111$>$ is relatively small \cite{Jeffries:13} and does not lead to latent heat formation in this material.

\section{Results and Discussions}

\begin{figure}[b]
\begin{center}
\includegraphics[angle=0, width=3.2 in]{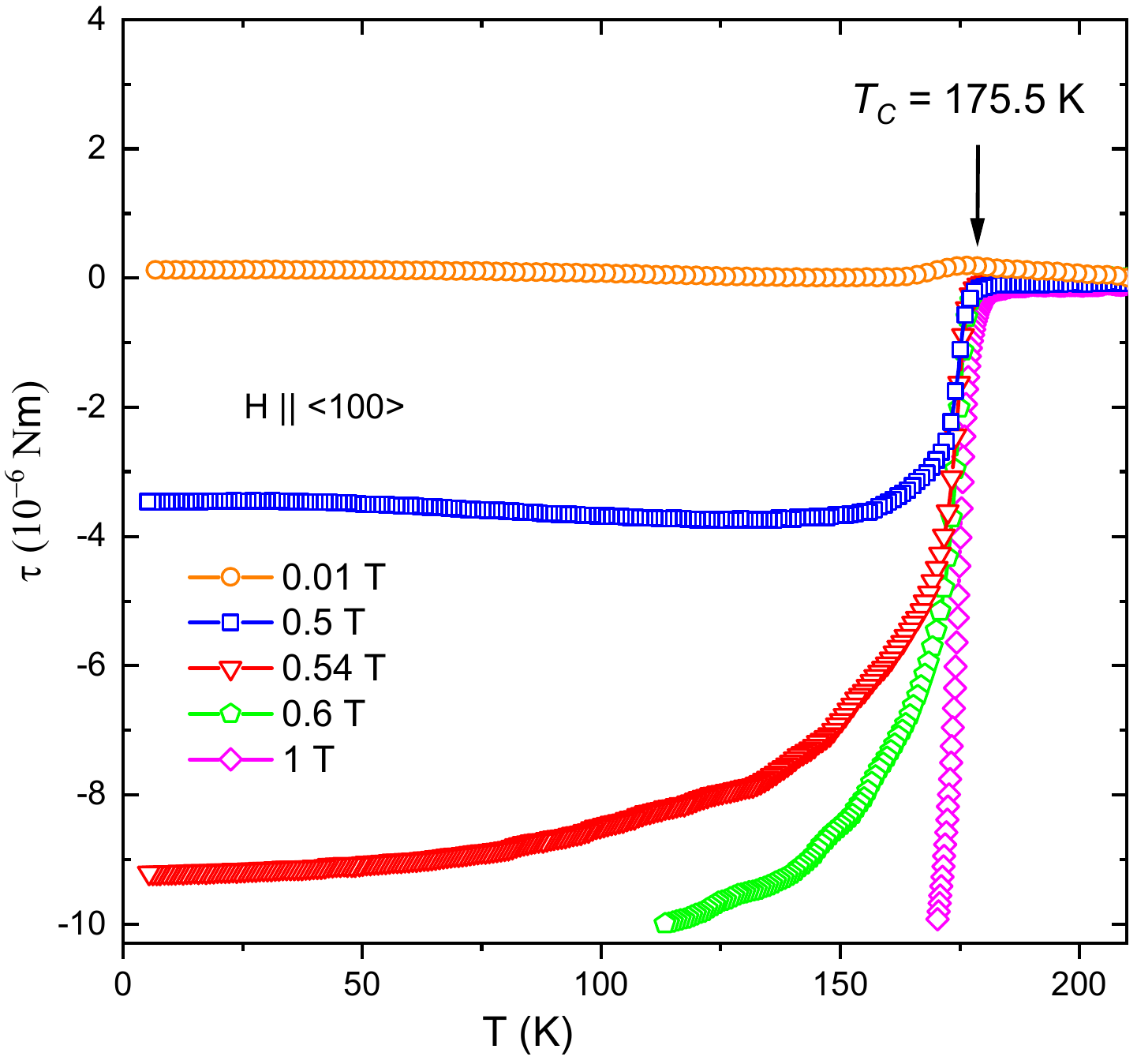}
\end{center}
\caption{(Color online) The temperature dependence of magnetic torque measured at different magnetic fields. The magnetic field is applied parallel to $<$100$>$ direction.}
\label{fig1}
\end{figure}

The figure \ref{fig1} shows the temperature dependence of the magnetic torque of US crystals in PM and FM phases and measured at various magnetic fields. All the measurements have been performed with a magnetic field applied along the $<$100$>$ crystallographic direction of the crystal. The $<$100$>$ direction has been shown to be the hard axis in US \cite{Tillwick:76b} and the largest torque is expected in this direction. In general, in external magnetic field a crystal experiences a magnetic torque, if there is any anisotropy of the magnetic susceptibility and the field is not applied along principal axes. The torque $\tau$ is defined by the following relation:
\begin{equation}
\overrightarrow{\tau}  =\overrightarrow{M} \times \overrightarrow{H},
\label{eq1}
\end{equation}

\begin{figure}[b]
\begin{center}
\includegraphics[angle=0, width=3.2 in]{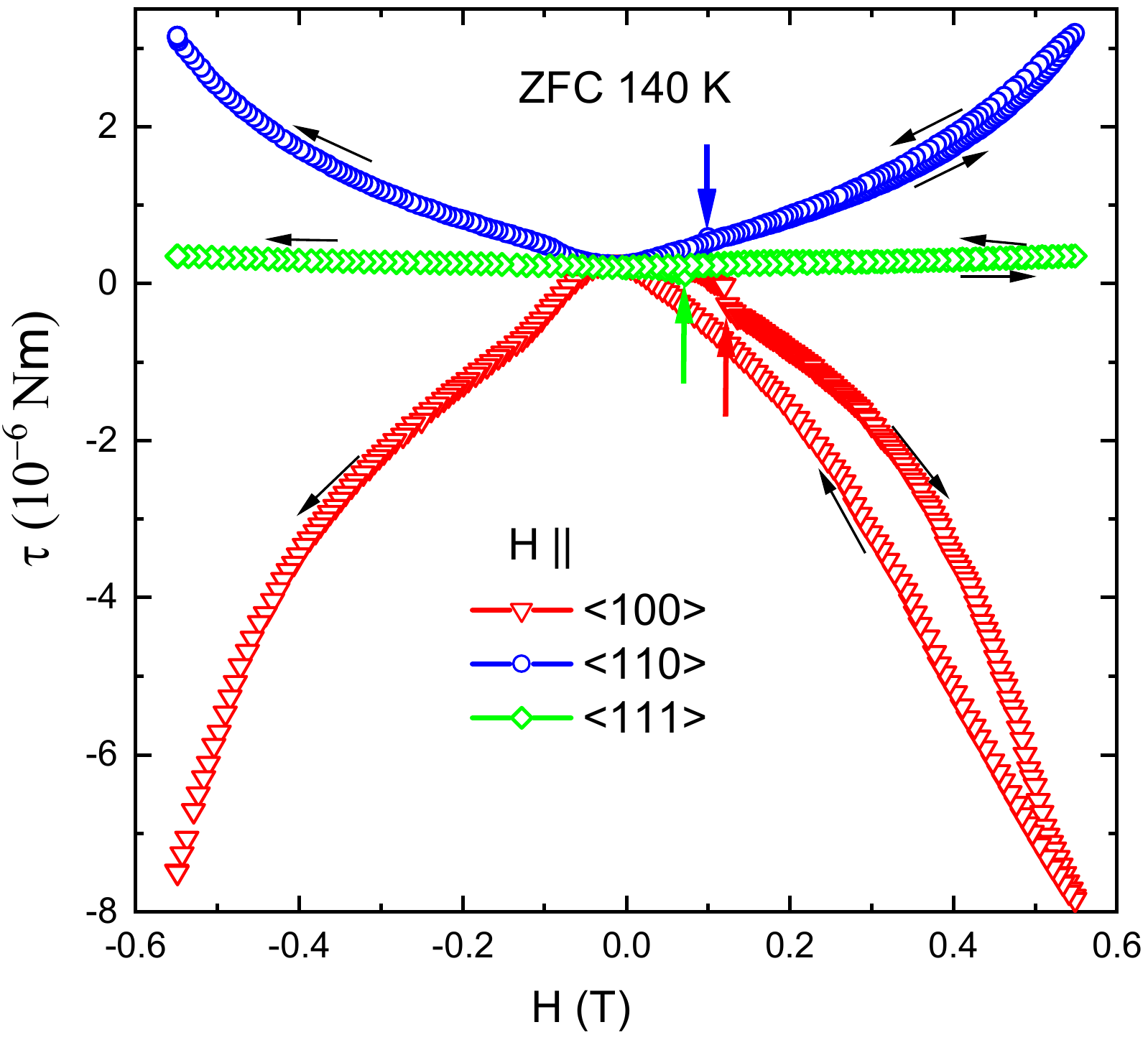}
\end{center}
\caption{(Color online) The magnetic field dependence of magnetic torque along the three principal directions of US single crystal measured in the FM phase (T =140 K). The $<$100$>$ direction is the hard axis in the ferromagnetic phase and maximum torque is expected in this direction.} The black arrows indicate the data taken for increasing and decreasing fields. The field was increased from zero to maximum value, then decreased to zero and again increased towards negative direction. The red, blue, and green arrows mark the change in the domain structure of the US crystal (see the text).
\label{fig2}
\end{figure}

where $\overrightarrow{M}$ is the magnetization of a sample and $\overrightarrow{H}$ is applied magnetic field, and therefore the torque is a measure of the perpendicular component of the crystal magnetization. As shown in Fig. \ref{fig1}, the PM to FM transition in US is clearly marked by a large drop towards large negative $\tau$ values, when the sample enters into the FM phase at 175.5 K. The torque magnitude in 0.54 T and at 5 K is close to -9.2$\times$10$^{-6}$ Nm. For higher magnetic fields, the torque response is so high that it crosses the limit of the torque cantilever (10$^{-5}$ Nm). This unfortunately limits the magnetic field that can be used in our experiments to fields below the saturation field $\sim$2 T \cite{Tillwick:76}. In the paramagnetic range, the torque is very small and lies within the background values for magnetic fields up to 1 T. As expected, the torque magnitude suddenly increases in the FM phase, and saturates at low temperatures. Fig. \ref{fig2} shows the magnetic torque scans versus magnetic fields for three main crystallographic directions at 140 K. Before each field scan, the crystal was cooled in a zero magnetic field. The black arrows in the figure indicate the data taken with increasing and decreasing magnetic field. At first, the field was increased from zero to a maximum value (to avoid crossing 10$^{-5}$ Nm), then decreased to zero and again increased towards a negative direction. As might be seen from the figure, the magnitude of the torque is the largest along $<$100$>$ (hard axis) and smallest along $<$111$>$ (easy axis) crystallographic directions. The vertical arrows in Fig. \ref{fig2} marked the sudden change in torque value in the initial increase of the field. Similar anomalies have been observed in low field magnetization measurements and have been attributed to a change in magnetic domain structure in this material \cite{Tillwick:76, Tillwick:77}.



\begin{figure}
\begin{center}
\includegraphics[angle=0, width=3 in]{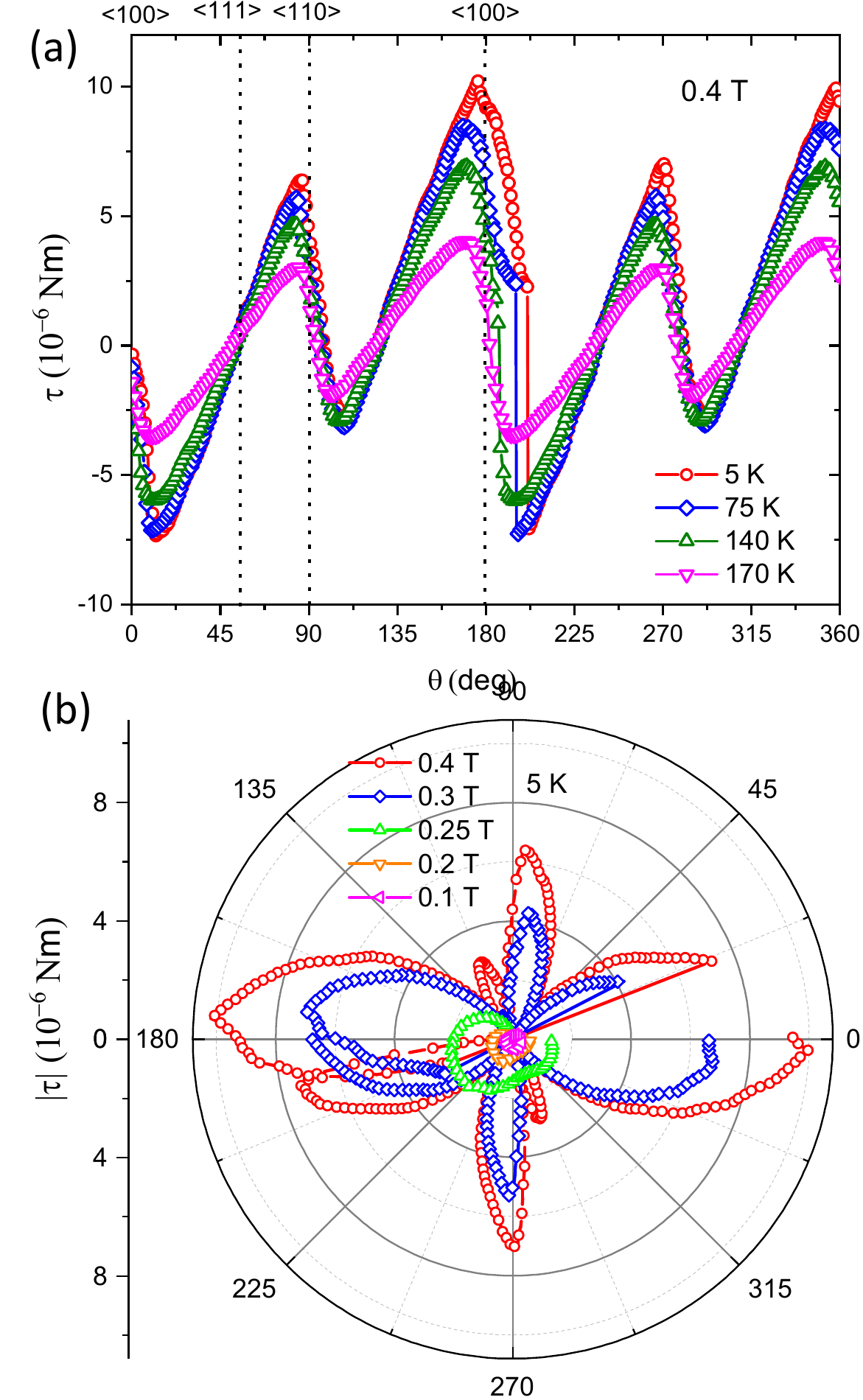}
\end{center}
\caption{(Color online)  (a) The angular dependence of magnetic torque measured at various temperatures in the FM phase (H = 0.4 T). (b) Polar plot of the magnetic torque of US crystal taken at different fields (T = 5 K).}
\label{fig3}
\end{figure}

The angular dependence of magnetic torque in the FM phase at different temperatures and at the magnetic field of 0.4 T is shown in Fig. \ref{fig3}(a). All the measurements were performed by cooling the sample in zero fields to the designated temperature, applying the magnetic field along with $<$100$>$ orientation, and then rotating the crystal. By rotating the sample in the horizontal rotator from $0^{\circ}$ to $360^{\circ}$, as shown in the inset of Fig. \ref{fig0}(a), one can align the magnetic field to all three principal directions of a cubic crystal; $<$100$>$, $<$111$>$, and $<$110$>$ at $0^{\circ}$, $54.7^{\circ}$, and $90^{\circ}$, respectively. The different orientations are marked by vertical dashed lines in Fig. \ref{fig3}(a). At low temperature, a sharp transition from positive to negative values is observed during the rotation. This is because of a sudden change in orientation of spins relative to the applied magnetic field when the sample is rotated. From the definition of the magnetic torque (Eq. \ref{eq1}), there shouldn't be any magnetic torque if the field is parallel to the easy axis, while maximum torque is expected if the field is parallel to the hard axis. At $54^{\circ}$, where the field is parallel to the easy axis $<$111$>$  \cite{Tillwick:76b}, the torque is close to zero as expected. On the other hand, the torque shows a maximum value at $180^{\circ}$, when the field is parallel to the hard axis $<$100$>$ \cite{Tillwick:76b}. At $90^{\circ}$, the field is parallel to $<$110$>$ which is the medium axis, and the torque magnitude is in between the values of easy and hard axes. 

The angular dependence of the zero-field cooled magnetic torque in the FM phase measured at different magnetic fields is presented in Fig. \ref{fig3}(b). At lower magnetic field values (up to 0.2 T), the torque is similar to the paramagnetic torque observed at low field. This indicates that the magnetic field is not sufficient to orient magnetic domains in the zero-field cooled sample. At 0.25 T, the torque shows slightly anisotropic behavior and above 0.25 T a large torque is observed with a similar anisotropy as seen in Fig. \ref{fig3}(a).

The free energy $F_a$ associated with crystalline anisotropy for a cubic crystal can be written in ascending powers of $\alpha's$ (direction cosines of the magnetization vector referred to the cubic axes of the crystal);
\begin{equation}
F_a = K_0+K_1(\alpha_1{^2}\alpha_2{^2} +\alpha_2{^2}\alpha_3{^2}+\alpha_3{^2}\alpha_1{^2})+K_2\alpha_1{^2}\alpha_2{^2}\alpha_3{^2}+...,
\label{eq2}
\end{equation}

where $K's$ represent the ferromagnetic anisotropy parameters. The magnetic torque can be analyzed using Fourier series as in \cite{Delmoral:87, Michelutti:92}
\begin{equation}
\tau (\theta) = Q_o + \sum_{n=1}^{\infty} Q_n sin n\theta,
\label{eq3}
\end{equation}

where $Q^{'s}$ are the torque amplitudes. The anisotropy constants can be expressed in terms of torque amplitudes as $K_1 = -\frac{8}{5}(4Q_2-Q_4)$ and $K_2 = -\frac{125}{8} (\frac{3}{2}Q_2-Q_4)$. The value of  $Q^{'s}$ depend on magnetic field at any temperature. In US, the preferential magnetization directions are parallel to the 111-type cube diagonals and the energy landscape is known as nickel-type anisotropy \cite{Busch:68}. This implies a negative $K_1$ parameter and the $K_2$ anisotropy parameter would range from -$\infty$ to 9 $\mid K_1\mid$/4 \cite{Cullity:08}. The torque measurements above the saturation field have to be conducted to calculate the $K_1$, $K_2$ precisely. In our measurements, the maximum field applied in FM phase is 0.4 T, which is lower than saturation field \cite{Tillwick:76}.

\begin{figure}
\begin{center}
\includegraphics[angle=0, width= 3.2 in]{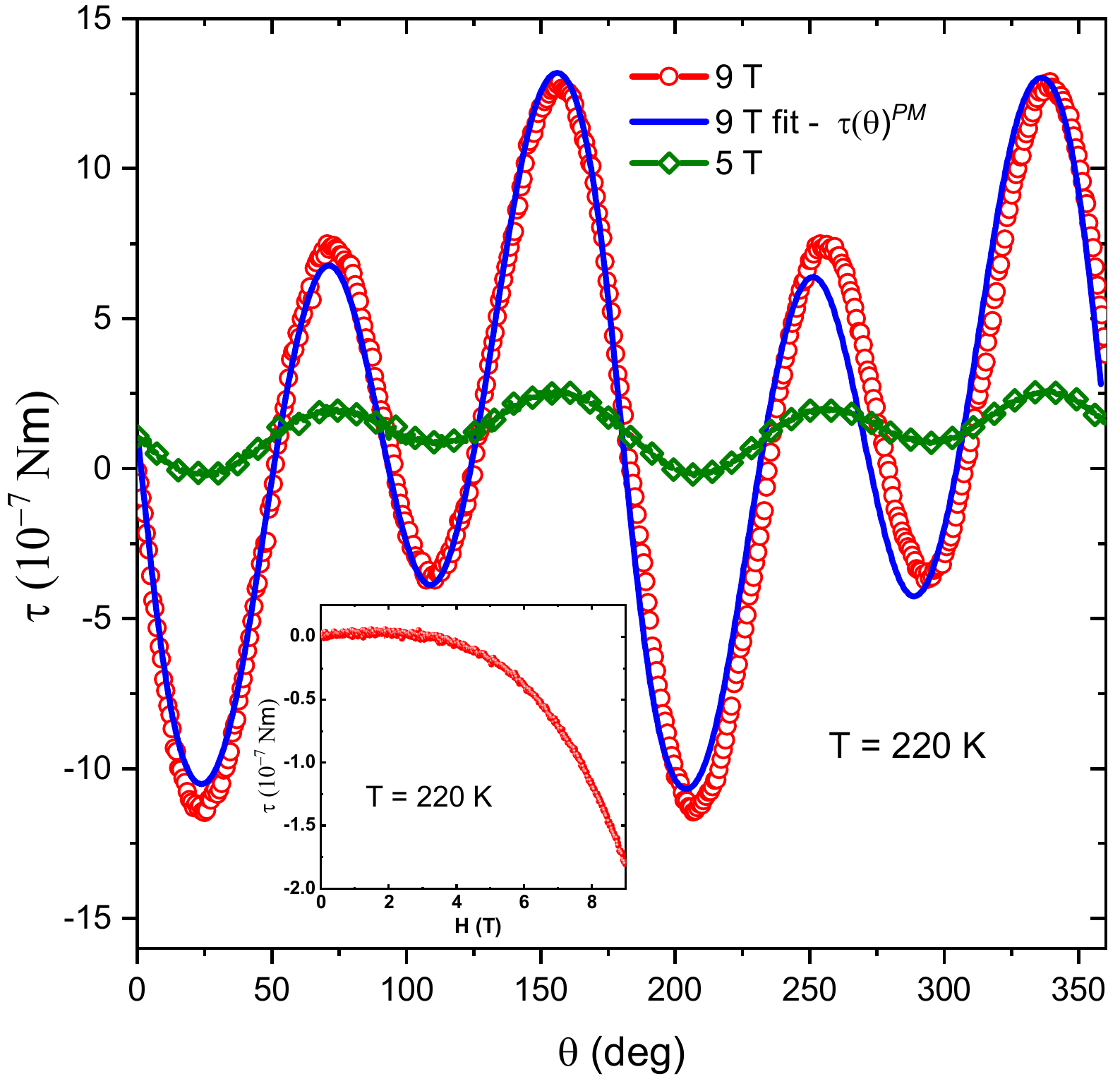}
\end{center}
\caption{(Color online) The angle dependence of the magnetic torque measured at 220 K and in the magnetic fields of 5 and 9 T.  The blue curve represents the fitting of Fourier series $\tau(\theta)^{PM}$ to the 9 T data. The crystal was rotated in two-fold plane \{110\}. The inset shows the field dependence of magnetic torque at 220 K when measured along the  $<$100$>$ direction.}
\label{fig4}
\end{figure}

The magnetic torque of US measured in the PM state is shown in Fig. \ref{fig4}. As can be seen from the figure, the angle dependence of the torque taken at different magnetic fields and at 220 K shows a two-fold symmetry. The maximum magnitude of the torque at 9 T is found to be 1.3$\times$10$^{-6}$ Nm, which is an order of magnitude larger than measured at 0.4 T.  Despite the weak signal, the two-fold rotational symmetry has also been observed at lower field values. The inset of Fig. \ref{fig4} shows the field dependence of magnetic torque along $<$100$>$ direction measured at 220 K. The field dependence of magnetic torque shows a characteristic quadratic behavior in agreement with the linear $M(H)$, observed in the paramagnetic state of US \cite{Tillwick:76}. 
The field dependence of magnetic torque shows a characteristic quadratic behavior in agreement with the linear M(H), observed in the paramagnetic state of US (cite M(H) results for US).

The magnetization of a crystal in paramagnetic phase can be expressed in terms of susceptibility tensors as:\cite{Delmoral:87}
\begin{equation}
\begin{split}
M_i =  \sum_{j=1}^{3}K_{ij}H_j + \sum_{j,k,l}\triangle_{ijkl}H_jH_kH_l\\
+\sum_{j,k,l,m,n}\Lambda_{ijklmn}H_jH_kH_lH_mH_n+...,
\label{eq4}
\end{split}
\end{equation}

In a perfect cubic crystal, the second rank susceptibility tensor $K_{ij}$ is isotropic and the other lowest order terms in $H$ giving anisotropy are the fourth and sixth rank susceptibility tensors $\triangle_{ijkl}$ and $\Lambda_{ijklmn}$, respectively. When the magnetic field is applied in the two-fold plane [110], the paramagnetic torque acting on the magnetization can be expressed as:\cite{Michelutti:92}
\begin{equation}
\tau (\theta)^{[110]}  = \frac{\chi_{||} - 3\chi_{\bot}}{8}[sin 2\theta +\frac{3}{2}sin  4\theta]H^4+ ...,
\label{eq5}
\end{equation}

where $\chi_{||} $ and $\chi_{\bot}$ are the fourth rank susceptibility tensors related to $\triangle_{ijkl}$ in Eq. \ref{eq4}. In cubic crystals, only even powers of H and even harmonics in $\theta$ are allowed, and the lowest order term producing the torque is proportional to $H^4$. If the crystal is not macroscopically cubic then $K_{ij}$ in Eq. \ref{eq4} is not isotropic and an additional term, $[(K_{||}- K_{\bot}) sin 2\theta]H^2$, which is proportional to $H^2$, arises in $\tau (\theta)$ of Eq. \ref{eq5} (see Ref. \onlinecite{Delmoral:87}). In both of the terms above, only a linear combination of the parallel and perpendicular susceptibilities can be obtained from the torque measurements in different crystallographic planes. The measured torque data at 9 T (Fig. \ref{fig4}) is analyzed by considering both $H^2$ and $H^4$ terms and neglecting higher-order terms. We also observed the non-zero torque amplitude corresponding to the $H^2$ term. That signifies that $K_{ij}$ is not isotropic, and the crystal is not macroscopically cubic in the PM phase. This behavior could be due to the anisotropic distribution of magnetic impurities and defects, which lowers the symmetry in the PM phase. The cubic crystals of DyAl$_2$ and ErAl$_2$ are also reported as macroscopically orthorhombic in the PM phase \cite{Delmoral:87} due to crystal imperfections. Other measurements such as thermal expansion and magnetostriction (parastriction) could help in revealing any macroscopic symmetry change in the US crystal.

\begin{figure}
\begin{center}
\includegraphics[angle=0, width=3.4in]{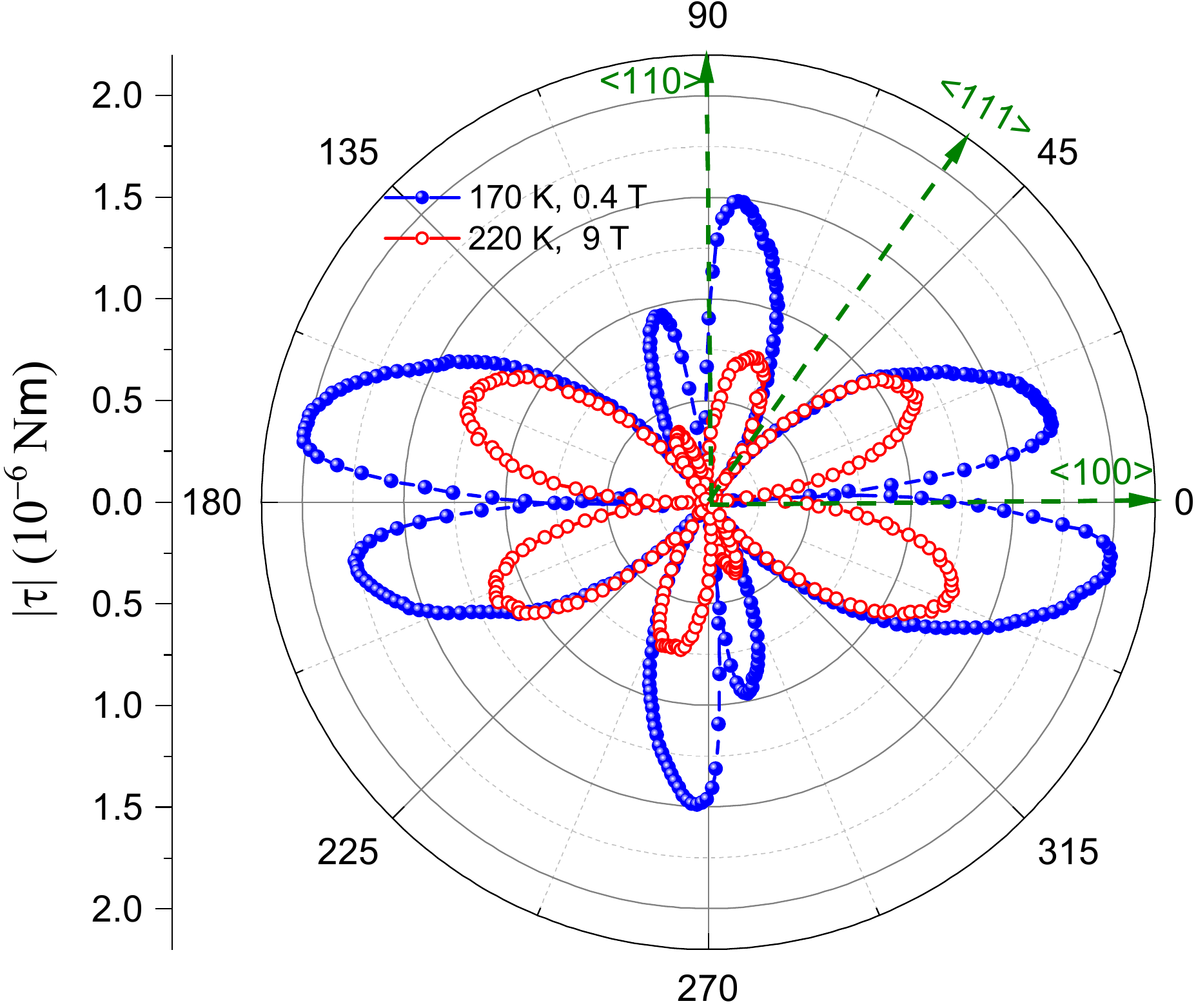}
\end{center}
\caption{(Color online) Polar plot of magnetic torque in paramagnetic (T = 220 K) and ferromagnetic (T = 170 K) phases of US crystal. The torque in the PM phase was measured at 9 T and FM torque was measured at 0.4 T. To compare the data qualitatively, the FM data was scaled down by factor of 1/2.}
\label{fig5}
\end{figure}

The observed magnetic torque in paramagnetic phase can also be analyzed using the Fourier series in Eq. \ref{eq3}. As US has cubic symmetry, only even harmonics of $\theta$ are allowed. However,  $n=1$ is also considered for the systematic error correction. The torque data for 9 T shown in Fig. \ref{fig5} is fitted with  $\tau (\theta)^{PM} = Q_o +  Q_1$ $sin \theta + Q_2$ $sin 2\theta + Q_4$ $sin 4\theta + Q_6$ $sin 6\theta + .....$, up to the $n = 14$ terms. The $Q_{4}$ is found to be the largest value as expected. The ratio of  $Q_{4}/ Q_{2}$ is found to be $\approx$1.7. The similar ratio derived in Eq. \ref{eq5} is 1.5. The difference may be  due to the anisotropic contribution from the $2\theta$ term. $Q_{6}$ is $\approx$2.5\% of $Q_4$ and $Q_{14}$ is less than 1\% of $Q_4$. The error correction term $Q_1$ is also small (2.3\% of $Q_4$). Based on this analysis, the paramagnetic torque in US can be explained from the fourth rank and second rank susceptibility terms in Eq. \ref{eq4}. In previous studies, the paramagnetic torque in 4$f$ electron systems of cubic crystals TmCd, TmZn and PrPb$_3$ were explained by fourth rank susceptibility tensor which is related to the quadrupole pair interactions \cite{Michelutti:92, Morin:82}. In addition, the symmetry lowering corresponding to quadrupolar order parameter was also mentioned in these systems. A symmetry lowering effect as in \cite{Michelutti:92, Morin:82} may have been also present in US due to quadrupole interactions in 5$f$-electron systems related to fourth rank susceptibility induced by magneto-elastic effect.


The anisotropic behavior of US crystals in the FM phase is related to strong magneto-crystalline anisotropy and spin-orbit coupling emerging from its 5$f$-electronic configuration. The observed magnetic anisotropy in the PM state might be due to the symmetry change from both anisotropic distribution of defects in the crystal and quadrupolar interactions \cite{Delmoral:84, Delmoral:87}. It would be interesting to compare qualitatively the anisotropic behaviors in these two phases. The magnetic torque data for PM and FM phases are shown together in the polar plot of Fig. \ref{fig5}. The torque in the FM phase was measured at 170 K (just below the $T_C$). As can be seen from the figure, the angle dependence of the torque in the PM phase has a similar type of anisotropy as observed in the FM phase. We have analyzed the $\tau(\theta)$ in the FM phase using Fourier series (Eq. \ref{eq3}) and observed that the $Q_4$ parameter is has the largest torque amplitude, as in PM phase. Furthemore, the $Q_2$ parameter accounts for 44\% of $Q_4$ and $Q_6$ stands for 10\% of $Q_4$, which are both a larger percentage than observed in the PM phase. Therefore, the anisotropy observed in FM phase can also be described by the fourth and second rank susceptibilities as in the PM phase. The similar type of anisotropy observed below and above $T_C$ was also reported in a ferromagnetic crystal, TbAl$_2$, where the anisotropy was caused by defects and higher rank paramagnetic susceptibilities \cite{Delmoral:84}. In the case of uranium monosulfide, it has been reported that the crystal undergoes a rhombohedral distortion along [111] crystallographic direction \cite{Wedgwood:72, Lander:90, Herrm:06} below $T_C$. Our torque measurements suggest that, in the PM phase, the applied magnetic field may have induced a similar distortion as observed in the FM phase. More detailed crystallographic studies under the magnetic field would be necessary to confirm the hypothesis, and to better understand the anisotropic behaviors.

\section{Summary and conclusions}

In summary, we have studied the magnetic anisotropy in uranium monosulfide single crystals by performing detailed magnetic torque measurements. It crystallizes in the cubic rock salt type of crystal structure and exhibits one of the largest magneto-crystalline anisotropy observed in cubic systems. We show that in the paramagnetic phase both second-rank and fourth-rank magnetic susceptibilities contribute to the observed anisotropy of US by reducing the cubic symmetry. As expected, we also observe a large magnetic anisotropy in the ferromagnetic phase with the $<$100$>$ direction being a hard axis, $<$111$>$ represents easy axis, and $<$110$>$ is an intermediate axis (saddle point). Furthermore, the similar anisotropic behaviors observed in both phases (PM and FM) suggests a presence of a similar type of crystallographic distortion (presumably rhombohedral) induced by internal or applied magnetic fields. Further study of the magnetostriction (parastriction in particular) and x-ray diffraction in the magnetic field will help to unveil the details of the magneto-structural phase transition and distortion in this material. Furthermore, the magnetic torque measurements could also be expanded further to better understand the magnetism and magnetoelastic interactions in other $f$-electron systems.

\begin{acknowledgments}
This work was supported by  the Idaho National Laboratory's Laboratory Directed Research and Development (LDRD) program (18P37-008FP) and DOE's Early Career Research Program. Portions of this work were performed under the auspices of the U.S. Department of Energy by Lawrence Livermore National Laboratory under Contract DE-AC52-07NA27344.

\end{acknowledgments}

\bibliography{US}

\end{document}